\documentclass{elsart}
\usepackage{epsfig}

\begin{document}
\begin{frontmatter}
\title{Mars Express/ASPERA-3/NPI and IMAGE/LENA observations 
of energetic neutral atoms in Earth and Mars orbit}

\author{M. Holmstr\"om${}^{a}$, M.R. Collier${}^{b}$, S. Barabash${}^{a}$,}
\author{K. Brinkfeldt${}^{a}$, T.E. Moore${}^{b}$, and D. Simpson${}^{b}$}



\address[Sweden]{Swedish Institute of Space Physics, Box 812, S-98 128, Kiruna, Sweden, e-mails: matsh@irf.se, stas@irf.se, klas@irf.se}
\address[GSFC]{NASA/GSFC Code 673, Greenbelt, MD. 20771, e-mails: Michael.R.Collier@nasa.gov, Thomas.E.Moore@nasa.gov, David.G.Simpson@nasa.gov}

\begin{abstract}
COSPAR Paper Number: D1.1-0088-06

\noindent
The low energy neutral atom imagers on Mars Express and IMAGE have revealed that the neutral atom populations in interplanetary space come from a variety of sources and challenge our current understanding of heliospheric physics. For example, both in cruise phase and at Mars, the neutral particle instrument NPD on Mars Express observed ``unexplained neutral beams" unrelated to Mars which appear to be either of heliospheric or solar wind origin. Likewise, the NPI instrument on Mars Express has revealed streams of neutral atoms with different properties than those observed by NPD. Independently, IMAGE/LENA has reported neutral atom observations that may be interpreted as a ``secondary stream" having different characteristics and flowing from a higher ecliptic longitude than the nominal upstream direction. Both sets of observations do not appear to fit in easily with the neutral atom environment from 1.0-1.57~AU as it is currently understood. In this paper we examine some highly suggestive similarities in the IMAGE/LENA and Mars Express/ASPERA-3/NPI data to try to determine potential origins for the observed signal.
\end{abstract}

\end{frontmatter}

\section{Introduction}

\begin{figure}
\begin{center}
\epsfig{file=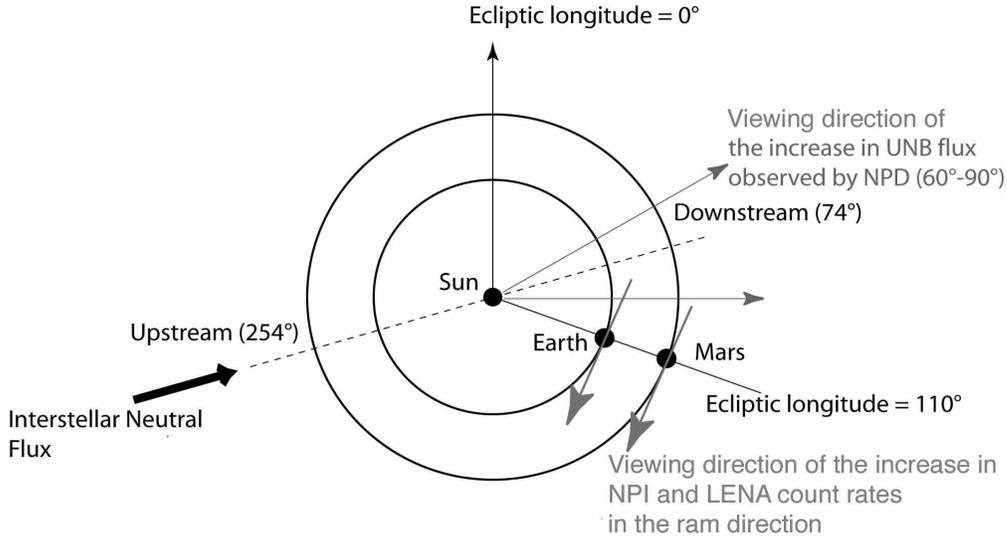,height=2.8in,width=5.2in}
\end{center}
\caption{Schematic showing NPI and LENA viewing directions along with
the viewing directions where ASPERA-3/NPD observed increases in the flux of
``Unknown Neutral Beams" [Galli et al., 2006]. Aside from all observations being in the nominal downstream region, there does not appear to be any correlation between the NPI/LENA and the NPD observations}
\end{figure}

\ \ \ Until recently, it was believed that the energetic neutral atom environment in the heliosphere results from the motion of the heliosphere at about 25~km/s through the local interstellar cloud. This relative motion creates an apparent stream of neutral atoms which approaches the Sun from the ``upstream" direction and flows away from the Sun toward the ``downstream" direction. When the Earth sits at 74${}^\circ$ ecliptic longitude in early June of every year, it is upstream of the Sun while six months later in early December it sits at 254${}^\circ$ ecliptic longitude, downstream of the Sun. This flow was expected to produce a neutral particle population that exhibits symmetry with respect to the 74${}^\circ$/254${}^\circ$ ecliptic longitude axis, the direction of relative motion. 
However, an increasing number of data sets related to neutral atoms do not conform to this expectation [e.g. Collier et al., 2004;
Wurz et al., 2004; Galli et al., 2006], suggesting that the heliospheric neutral atom environment may be more complex than previously thought.
In this paper, we will address two of the neutral atom data sets that do not conform to expectations: one from the Low Energy Neutral Atom (LENA) Imager and one from
the Mars Express/ASPERA-3/Neutral Particle Imager (NPI), both low energy neutral
atom sensors.

\ \ \ Figure~1 shows the heliospheric context of these neutral atom observations and summarizes the Mars Express NPI, Mars Express NPD [Galli et al., 2006] and IMAGE/LENA [Collier et al., 2004] neutral atom observations.
The interstellar neutral flux arrives from about 254${}^\circ$ ecliptic longitude and close to the ecliptic plane [e.g. Frisch, 2000] and is represented by the arrow on the left. The NPD observations appear to come from practically every ecliptic longitude and have been interpreted as $\sim$1~keV hydrogen. However, the highest flux 30${}^\circ$ bin falls between 60${}^\circ$ and 90${}^\circ$ ecliptic longitude, although with relatively few events (see Galli et al., Figure~9). This viewing direction implies a neutral atom flow preference {\it antiparallel} to the interstellar neutral flux direction.

\ \ \ The IMAGE/LENA and Mars Express NPI observations both occur when the observation point itself (i.e. Earth or Mars, respectively) is at about 110${}^\circ$ ecliptic longitude. At this ecliptic longitude, the NPI data are observed flowing in the Mars ram direction and the LENA observations are consistent with flow in the Earth ram direction, as indicated by the arrows. By ``ram direction" we refer to the direction from which a stationary or nearly-stationary particle would be observed to come due to the motion of the planet. This direction is roughly perpendicular to the direction of the Sun.
Finally, as will be shown, the shape of the count rate profile observed by LENA
is similar to that observed by NPI.

\ \ \ The peak fluxes for all three instruments are observed in the nominal downstream direction, but there does not exist any correlation between the direction of the interstellar flow and the direction of the peak fluxes, which arrive from the opposite hemisphere.

\section{Mission Descriptions}

\ \ \ Mars Express, launched in June of 2003, was both the first ESA mission to Mars at 1.57~AU and the first European mission to any other planet. The spacecraft was captured into Mars orbit in December 2003 with spacecraft commissioning ending in mid-January 2004 and the instruments on the orbiter beginning commissioning thereafter. The Mars Express orbiter scientific payload includes seven experiments, one of which is the ASPERA Energetic Neutral Atoms Analyzer [Chicarro et al., 2004].

\ \ \ Mars Express is a 3-axis stabilized orbiter with a fixed high-gain antenna and body-mounted instruments in an elliptical, quasi-polar orbit 250 by 10,142~km [Chicarro et al., 2004]. The NPI's 360${}^\circ$ by 9${}^\circ$ aperture is split into 32 sectors numbered 0 to 31, each 11.25${}^\circ$ by 9${}^\circ$. All sector view directions are equally spaced within the YZ plane of the MEX/ASPERA frame. The boundary between sectors 0 and 31 is coincident with the frame's Z-axis so that the 0 sector view direction is 5.625${}^\circ$ off the +Z axis towards +Y.

\ \ \ The Imager for Magnetopause-to-Aurora Global Exploration (IMAGE) mission was the first satellite dedicated to imaging the Earth's magnetosphere using ultraviolet imaging, energetic neutral atom imaging and radio plasma imaging. One of the neutral atom imagers on this satellite is the Low Energy Neutral Atom (LENA) imager, designed to explore Earth's neutral atom environment at energies as low as 10 eV [Burch, 2000].

\ \ \ The IMAGE observatory is spin-stabilized and oriented so that the IMAGE viewing instruments scan the Earth each observatory revolution, about two minutes, from its highly elliptical 1000~km x 7~R${}_{\rm E}$ altitude [Gibson et al., 2000]. As the spacecraft rotates, the LENA data are collected in forty-five 8${}^\circ$ sectors. The start of sector 0 is spacecraft nadir and LENA observes the Earth at 135${}^\circ$ or close to the boundary between sectors 16 and 17.

\section{Instrument Descriptions}

\ \ \ Both LENA and NPI employ surfaces to detect low energy neutral atoms. LENA uses the surface to convert the neutral atoms, which interact with the surface at a grazing incident angle of 15${}^\circ$, to negative ions. Following conversion, the negative ions are accelerated by an extraction lens to about 12~kV, traverse an hemispherical 
electrostatic deflection system, and enter a time-of-flight/position 
sensing subassembly which measures species and polar angle. Spacecraft 
spin determines azimuth angle with the two minute IMAGE spin divided up 
into 45~sectors each of 8${}^\circ$ angular width [Moore et al., 2000]. 

\ \ \ To determine the mass of the incident low energy neutral atoms, the LENA instrument includes a time-of-flight section. The time-of-flight of the ion entering this subsection, which arrives at a predetermined energy because the conversion surface is at a large negative voltage (typically -12~kV), is determined by the delay between two signals, the start signal generated by secondary electrons when the ion passes through a thin carbon foil, and a stop signal generated by the ion itself hitting a microchannel plate. Because the efficiency for generating each of these signals is less than unity, the ``coincidence rate" is considerably lower than the rate for the individual start and stop signals, the ``singles rates." Consequently, low rate signals, such as the one described here, are easier to see in the singles rates.

\ \ \ The NPI sensor is similar to those developed for ASPERA-C/Mars-96 and Astrid. As in LENA, the charged particles are removed by an electrostatic deflection system which also collimates the incoming beam.
The region between the deflection system is divided up into 32 sectors by
plastic spokes which form 32 azimuthal collimators each having an aperture
of 9${}^{\circ}$x18${}^{\circ}$. Neutral atoms pass through the deflection system and interact with a conical target at a grazing incidence angle of 20${}^\circ$.
This interaction generates secondary particles and/or the reflection of the
primary neutral atom. A microchannel plate stack then detects the particles
leaving the surface [Barabash et al., 2004].
In the sense that a neutral atom needs to trigger only one detector stack
to register a count, the NPI data are also a ``singles rate."

\ \ \ The MEX/NPI calibration shows a sharp low energy cut-off in the sensor efficiency at about 200~eV, the lowest energy NPI can observe (based on an H${}_2$O${}^+$ beam) [Holmstr\"om et al., 2006, Figure 7]. IMAGE/LENA calibration shows a sharp cutoff in the atomic hydrogen response between 15 and 20~eV [Collier et al., 2001, Table 1]. The difference in response at the low energies is due to the difference in technique: MEX/NPI records a count whenever the particle's incident energy is sufficient to trigger the microchannel plate stack. LENA only records a count if the neutral atom charge exchanges at the conversion surface from which it is accelerated into the analyzer.

\ \ \ One of the complications in analyzing the MEX/NPI data is that the sensitivity and background noise levels of the various sectors are different. Consequently, the MEX data from different sectors in this paper are sometimes shown shifted relative to each other [see, for example, Holmstr\"om et al., 2006].

\section{LENA Observations}

\begin{figure}
\begin{center}
\epsfig{file=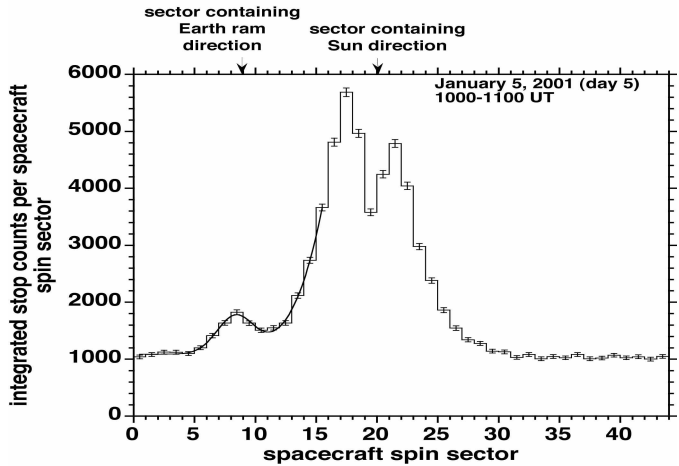,height=2.5in,width=3.5in}
\end{center}
\caption{One-hour averaged spin spectrogram in early January 2001 showing the presence of the annual downstream signal in the LENA data.}
\end{figure}

\ \ \ Figure~2 shows a LENA spin plot of a singles rate, the stop count rate, averaged over an hour from January 5, 2001. Because Fig.~2 shows a singles rate
and each sector includes a 45${}^\circ$ polar field-of-view, the broad peak located around the arrow indicating ``sector containing 
Sun direction'' is primarily the microchannel plate response to solar photons. The smaller peak, whose location is consistent with the Earth ram direction, is a signal observed annually by
LENA, but only in the downstream region. 
Note that this peak is about 90${}^\circ$ from the Sun direction.
Also, as shown in Figure~3, this signal has its peak flux at a consistently higher ecliptic longitude, about 112${}^\circ$, than expected based on the nominal upstream direction, 74${}^\circ$. This signal is described in Collier et al. [2004], and Figure~1 of that paper shows this signal in spectrogram format.
The LENA observations and possible interpretations have been reported in the literature by Wurz et al. [2004] and by Collier et al. [2004]. 

\begin{figure}
\begin{center}
\epsfig{file=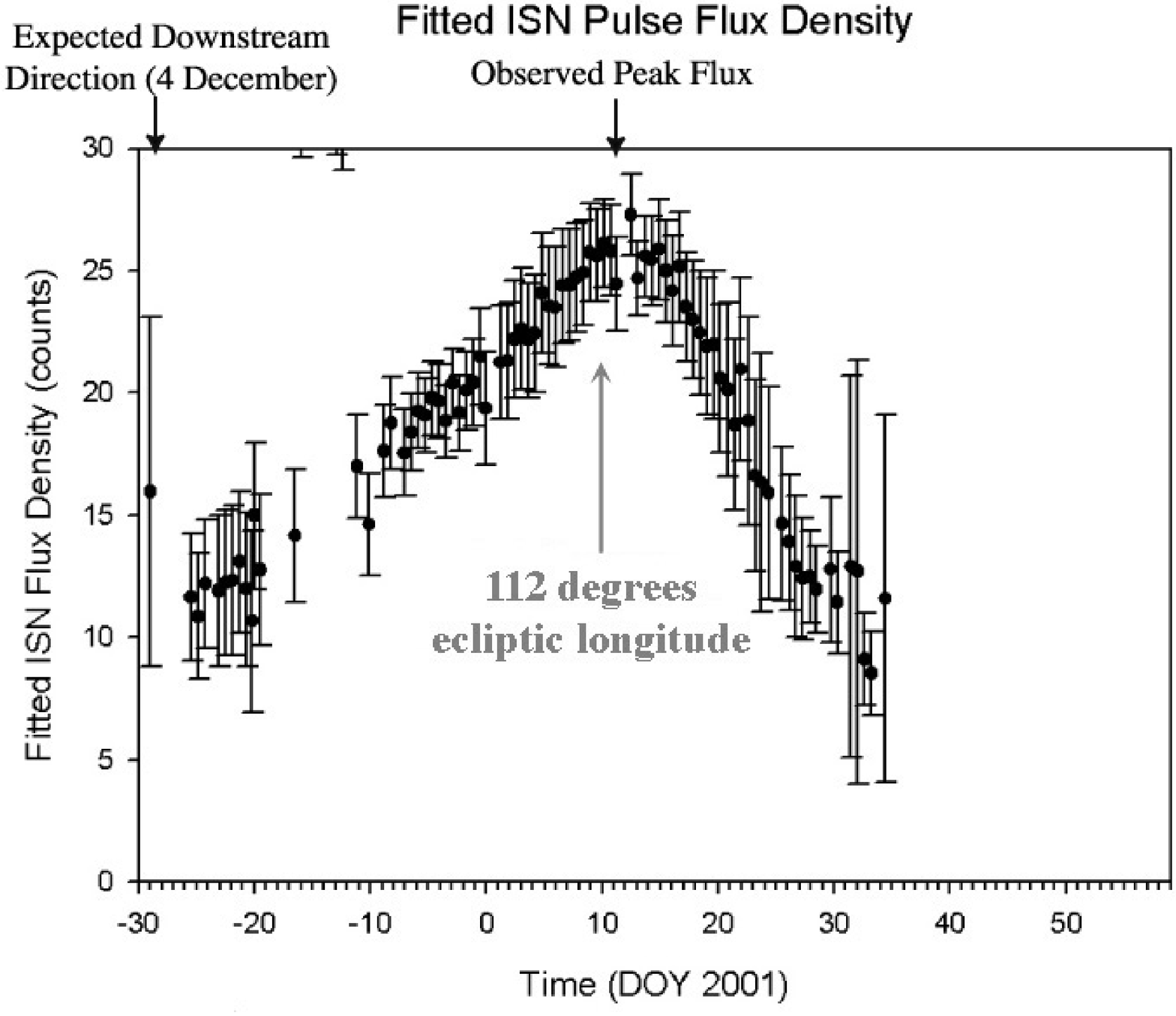,height=2.5in,width=2.5in}
\end{center}
\caption{The annual downstream signal observed by LENA peaks at about 112${}^{\circ}$ ecliptic longitude.}
\end{figure}

\section{NPI Observations}

\ \ \ As shown in Figures~2 and 3, the LENA signal occurs approximately in the Earth ram direction at about 110${}^\circ$ ecliptic longitude, or about 35${}^\circ$ higher in ecliptic longitude than the nominal downstream direction. Fortuituosly, Mars Express was in deep eclipse in the April-May time frame of 2004 while Mars was between 100${}^\circ$ and 110${}^\circ$ ecliptic longitude, approximately the same longitude where LENA sees the signal shown in
Figures~2 and 3.

\begin{figure}
\begin{center}
\epsfig{file=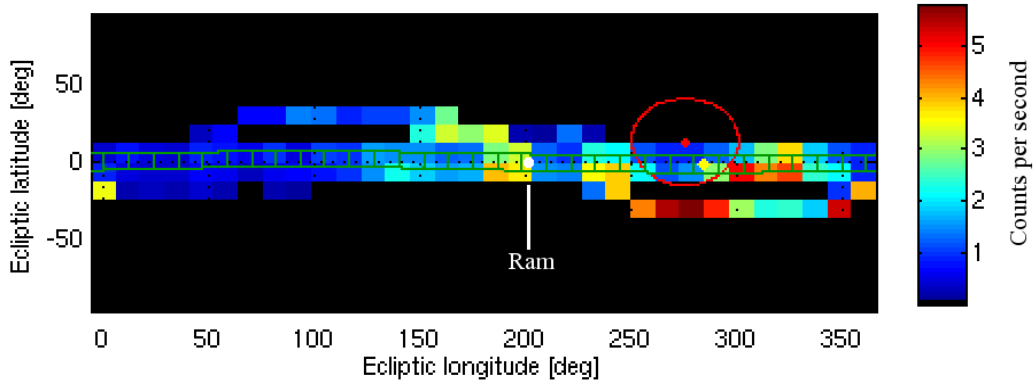,height=2.2in,width=6.0in}
\end{center}
\caption{NPI sky map generated as an average of observations in the shadow 
of Mars during April and May 2004 when Mars' ecliptic longitude was 100-110
degrees.  Black denotes no observations.
Indicated is Mars' ram direction. Mars, and its center, is shown in red.
The Sun direction is shown in yellow.  In green is illustrated the field of 
view of the NPI's 32 sectors at one time.}
\end{figure}

\ \ \ Figure~4 shows an NPI image from deep eclipse in Spring of 2004 when Mars stood between 100${}^\circ$ and 110${}^\circ$ ecliptic longitude. Plotted on the x-axis is the look direction ecliptic longitude and plotted on the y-axis is the look direction ecliptic latitude. During this period, the Sun, the nominal upstream direction, and the Galactic center direction were all occulted by Mars, shown by the red oval. One can see some signal on the limb of Mars which 
are neutral solar wind ENAs propagated into the Martian eclipse [Brinkfeld
et al., 2006]. Thus, this period is an ideal time to look for unexpected signals near the Mars ram direction.

\ \ \ Note that Figure~4 shows a number of bright pixels near 200${}^\circ$
ecliptic longitude, close to the Mars ram direction.
Thus, like LENA, ASPERA-3 observes a signal ninety degrees away from the Sun direction. So, ASPERA-3 may be seeing at Mars at 1.57~AU the same signal LENA observes at Earth at 1~AU.

\ \ \ One might think that this is simply the interstellar helium focusing cone.
However, the interstellar neutral helium energy observed at Mars is below the efficiency cutoff of the Mars Express NPI instrument. Furthermore, 
the direction of observation also suggests Mars Express is not observing the standard downstream He focusing cone, which ought
to be centered closer to downstream of 254${}^\circ$ ecliptic longitude, the nominal upstream direction. Instead,
the observations, at about 100${}^\circ$ ecliptic longitude, are nearer
in ecliptic longitude to
downstream of the Galactic center at 267${}^\circ$ and to downstream of
the apex of the Sun's
way, the motion of the Sun with respect to local stars, at about 271${}^\circ$
ecliptic longitude than to the nominal downstream direction 
of the interstellar flow at 74${}^\circ$.

\section{Ultraviolet Light}

\begin{figure}
\begin{center}
\epsfig{file=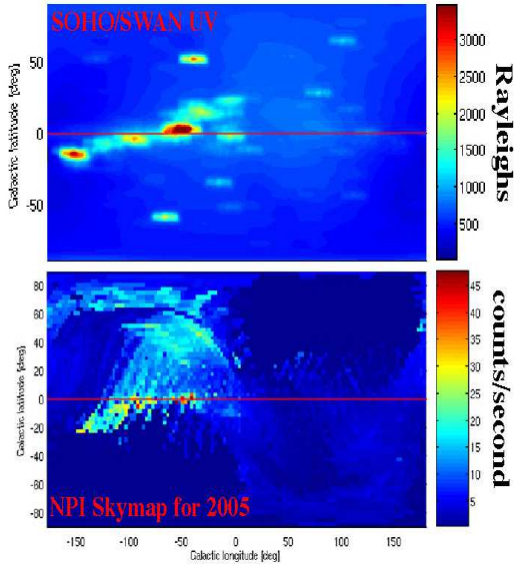,height=3.0in,width=3.0in}
\end{center}
\caption{Upper panel shows a SOHO/SWAN all-sky UV (117-180 nm) map in 
units of Rayleighs.
We have convolved the images with the response function of one NPI sector 
(from calibrations).
Data provided by Teemu Makinen on behalf of the SWAN team.
The lower panel shows an NPI sky map of average count rates [1/s] for eclipse
observations in 2005, i.e. we see the average count rate that the NPI sectors
have registered when looking in that direction.
For both panels the coordinate system is galactic with longitudes on the
x-axis and latitudes on the y-axes.}
\end{figure}

\ \ \ Figure~5 shows two images adapted from Holmstrom et al. [2006]. 
Here an all-sky UV map in Galactic coordinates adjusted for the NPI UV 
angular response from the SWAN instrument on SOHO is shown (Teemu Makinen gratefully acknowledged) along with NPI image count rates
in Galactic coordinates. 
The SOHO/SWAN image is one degree resolution in units of Rayleighs and
represents wavelengths between 117 and 180~nm. The image has been smoothed
with the NPI response function. The NPI data are from the eclipse observations
in 2005 and are average rates in counts per second. Note that the
data are in Galactic coordinates so that the Galactic plane is the red line
across the center of each plot.
In comparing the structure of the SOHO/SWAN data in the upper panel of Figure~5 with that of the MEX/NPI data in the lower panel, it appears that 1000 Rayleighs corresponds roughly to an NPI response of about 10 counts per second. Furthermore, the strong UV signals observed by SOHO near the Galactic plane between about -100${}^\circ$ and -50${}^\circ$ Galactic longitude appear to be mirrored in the NPI data in the lower panel, suggesting a strong UV response. However, if this were the whole story, we would observe a significant ($>$40 counts/s) NPI count rate slightly below the Galactic plane to the left of -150${}^\circ$ Galactic longitude based on the SOHO/SWAN data, as well as smaller isolated count rate enhancements near -50${}^\circ$ longitude, 50${}^\circ$ latitude and near -75${}^\circ$ longitude, -50${}^\circ$ latitude. None of these is observed in the NPI data 
although the signal just below the Galactic plane at -150${}^\circ$
longitude may be partly seen by NPI. Furthermore, some NPI data do not have
a counterpart in the UV observations, for example the signals near
Galactic latitude 70${}^\circ$ stretching from about longitude -180${}^\circ$
to about -50${}^\circ$, although there is no ram signal in 2005. Plots showing these data in ecliptic coordinates are available in Holmstr\"om et al. [2006].
The absence of some of the UV sources in the NPI observations is a puzzle,
but the data show that NPI is not responding exclusively to UV; 
the movement of signals with time offers further evidence.

\begin{figure}
\begin{center}
\epsfig{file=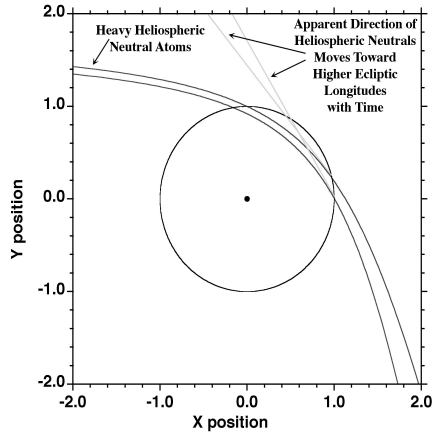,height=2.3in,width=2.3in}
\end{center}
\caption{Changes in the observed direction of a signal strongly suggests that signal is not UV emission from stars. For example, if the neutral atom trajectories  are
gravitationally bent by the Sun (at the origin), then the apparent direction 
of heliospheric neutrals shifts gradually to higher ecliptic longitudes with time if they are observed at an planet that orbits the Sun.}
\end{figure}

\ \ \ Because UV emissions from stars remain relatively stationary in fixed coordinates such as ecliptic or Galactic coordinates over the time-scale of ESA and NASA missions regardless of the position of observation in the inner
heliosphere, one way to deduce that an observed signal is not UV is to look for systematic changes in the observed direction with time. As an example,
if observations are made on a planetary orbit, the circle in Figure~6,
of neutral atoms, the hyperbolic trajectories, whose energies are low enough
for the Sun, at the origin, to gravitationally deflect their orbits, then the
direction of observation, shown by the lines, will shift to higher 
ecliptic longitudes systematically with time as the observation position orbits 
with the planet.

\ \ \ The signal observed in the ram direction, shown in Figure~4, drifts along with the Mars orbital motion. Figure~7 shows that the high longitude extent 
of this signal follows
the change of the Mars ram direction ecliptic longitude over 130~days,
labeled in white as ``upward drift" although the maximum intensity line does
not show as much drift. 
This behavior strongly suggests that this is not a UV signal. 
Also labeled in this figure are the positions of the Sun (yellow) and
the Earth (green).

\begin{figure}
\begin{center}
\epsfig{file=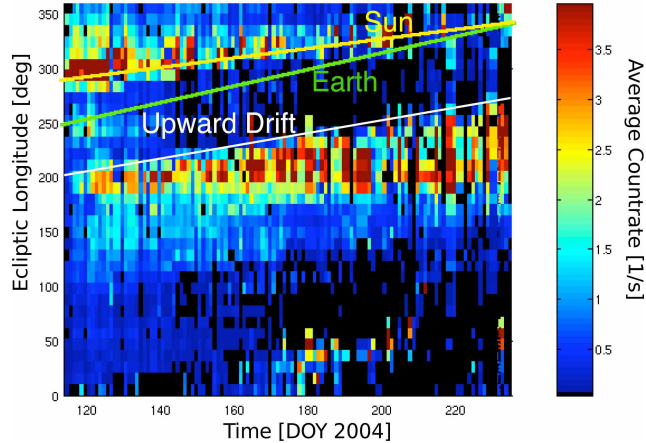,height=2.3in,width=3.3in}
\end{center}
\caption{Time evolution of the NPI signal in the Ecliptic plane during 2004 
is shown in white. 
The signal is from a direction near the ram direction and it shows a drift 
toward higher ecliptic longitudes with time.
Each column represent average count rates [1/s] during one day. The x-axis is
day of year 2004, and the y-axis is ecliptic longitude in degrees. 
Only observations by sectors looking at most 6 degrees off the ecliptic 
plane were included. We included all umbra observations. The green line
represent Earth's longitude, and the yellow line the Sun's longitude.}
\end{figure}

\section{MEX Upstream/Downstream Comparison}
\ \ \ If, in fact, this is a downstream phenomenon as the LENA data seem to suggest, then Mars Express NPI should observe a larger signal when looking in the same direction downstream near the ram direction versus upstream. Figure~8 compares data in the downstream from 2004 near the ram direction to that in the upstream region in 2005 when the same sectors were looking close to the same ecliptic longitudes. 
These sectors were viewing in the same direction because the spacecraft 
flipped over during the time period between the downstream and upstream
observations.
Sectors 30 and 31 which cover the ram direction during the 2004 downstream period reasonably well show a higher count rate than they do when looking at the same ecliptic longitude in the upstream region in 2005. 
Sector~30 shows a different shape and an enhancement downstream in the region in which it observes 190-200${}^\circ$ ecliptic longitude of over a factor
of three or so. Likewise, sector~31
over the region it observes 180-190${}^\circ$ ecliptic longitude shows a different shape and enhancement of a factor of two. Furthermore, the
signals in sectors~30 and 31 are similar in the downstream region.
Sectors~0 and 1 are higher in the upstream, but only by about 50\% in the
region of overlap of the observation directions and have the same general
trend in the upstream and downstream region suggesting that these sectors
may be observing the same signal, in contrast to sectors~30 and 31.
Any UV response would be approximately the same in 2004 and 2005. The fact that sectors~30 and 31 are enhanced in the downstream region and show
similar shapes suggests much or most of the signal is not UV.

\begin{figure}
\begin{center}
\epsfig{file=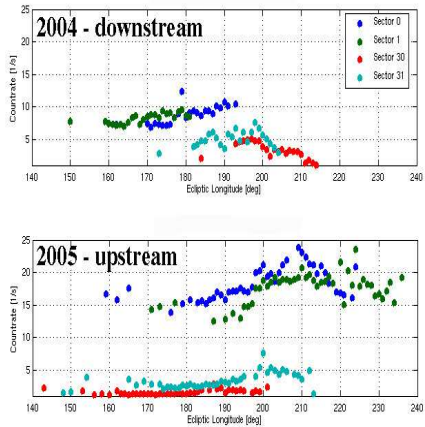,height=2.3in,width=2.3in}
\end{center}
\caption{Comparison of MEX data near the ram direction during eclipse
periods from 2004 to data in 2005. This figure only includes observations 
when the NPI field-of-view plane was in the ecliptic plane.
The x-axes are the ecliptic longitudes [deg] of the sector look directions,
and the y-axes are the average count rates [1/s] over the time periods.
We have averaged all umbra observations, but have not included any sector 
counts where the line of sight gets closer to Mars' surface than 1000~km.}
\end{figure}

\section{Comparison to LENA}

\ \ \ Although there may exist a Mars Express downstream signal at about the same
ecliptic longitude as the signal observed by LENA,
is there any evidence for a peak in the Mars Express data in ecliptic longitude
similar to the peak in the LENA data? Figure~9 shows, with the blue, black and red solid circles, the Mars Express data from 2004 in Figure~8 rescaled and plotted as a function of the observed ecliptic longitude for each sector. 
The observed ecliptic longitude for each sector is not exactly the ram
direction, but is close. For example, sector~30 observes its peak rate at
about 197${}^\circ$ ecliptic longitude when Mars is at 111${}^\circ$
ecliptic longitude. At this time, Mars' ram direction is 201${}^\circ$
ecliptic longitude.
The solid tan squares are the LENA data from Figure~3. Also shown as the top row of numbers above the upper x-axis is the ecliptic longitude of the Earth's ram direction for comparison to the Mars Express data. All three of the Mars Express sectors (but most notably 30 and 31) show evidence for a peak close to 200${}^\circ$ ecliptic longitude. 

\ \ \ The qualitative similarity of the time profiles of the Mars Express and LENA
observations suggests that the two instruments may be detecting the same
phenomenon, one at Earth at 1~AU and one at Mars at 1.57~AU.

\begin{figure}
\begin{center}
\epsfig{file=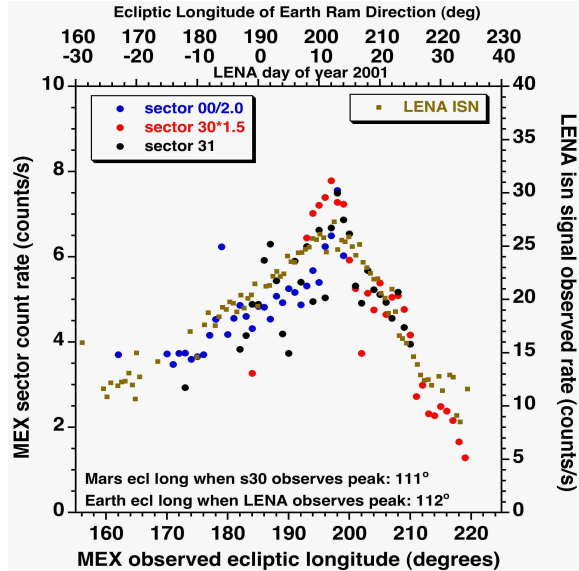,height=3.0in,width=3.0in}
\end{center}
\caption{Comparison between MEX/NPI data from 2004 (solid blue, red, and black circles) and LENA observations
(solid tan squares) downstream. This figure only includes observations when the NPI field-of-view plane was in the ecliptic plane. Note the similar shapes of the peaks: a slow rise and a faster decline with increasing observed ecliptic longitude.}
\end{figure}

\section{Conclusions}

\ \ \ The source of the signals reported here observed by IMAGE/LENA and Mars Express/NPI is unclear. The signals may be heliospheric in origin with the observation direction greatly influenced by the motion of Earth and Mars, respectively, in their orbits, as illustrated by Fig.~1. Whatever the origin, the IMAGE/LENA and Mars Express/NPI data show some striking similarities that do suggest the two instruments are observing the same signal.

\ \ \ First, the Mars Express signal is close to the Mars ram direction, as shown in Fig.~4, while the LENA data are consistent with being close to the Earth ram direction. This suggests the instruments are observing particles that are not moving with the planets but are being ``rammed" by the planets' motion. Signals associated with the planets themselves would not be expected to show this asymmetry.

\ \ \ Second, the two instruments observed the highest count rates from the ram direction when Mars and Earth had about the same ecliptic longitude, suggesting some sort of spatial structure exists at this longitude from at least 1~AU to 1.57~AU, similar to the downstream helium focusing cone. However, the observed signal does not appear to be associated with the interstellar neutral helium focusing cone, as the expected signal for the focusing cone is much smaller than the observed ``secondary stream" signal for LENA [Wurz et al., 2004] and the expected energy of the helium in the focusing cone falls below the Mars Express/NPI low energy cut-off.

\ \ \ Finally, the shape of the count rate profile in ecliptic longitude observed by LENA is similar to that observed by NPI (see Fig.~9). This might be expected if both spacecraft were traversing a spatial profile organized by longitude.

\ \ \ Thus, like many other data sets related to heliospheric neutral atoms, these two
neutral atom data sets do not exhibit the expected symmetry around the 74${}^\circ$/254${}^\circ$ axis. Like most of the other data sets, these
appear shifted toward higher ecliptic longitudes, suggesting some sort of heliospheric 
asymmetry. Indeed, Lallement et al. [2005], using SOHO/SWAN data have concluded that
the heliosphere is asymmetric with the heliopause closest to the Sun at higher ecliptic longitudes than the nominal upstream direction, perhaps as a result of
a tilted interstellar magnetic field [e.g. Ratkiewicz et al., 1998]. The Voyager~1 
LECP data also support the notion of a heliosphere substantially tilted toward higher
ecliptic longitudes than the nominal upstream direction [Decker et al., 2005].
Frisch [2006] has shown that small interstellar grains captured in interstellar
magnetic fields show a maximum polarization direction offset in ecliptic longitude 
about 35${}^\circ$ higher than the upwind direction, about the same magnitude of offset observed in the neutral atom observations reported here. She interprets these data as evidence for an interstellar magnetic field forming an angle of about 75${}^\circ$
with the nominal upstream direction. 
In short, the evidence does seem to suggest
that the symmetry axis of the heliosphere is not along the 74${}^\circ$/254${}^\circ$ axis but rather at higher ecliptic longitudes, closer
to the Galactic center direction or the apex of the Sun's way.

\ \ \ The field of low energy neutral atom imaging is very young and still maturing, relative to their charged particle cousins. The initial two low energy neutral atom imagers, the Swedish Astrid/PIPPI [e.g. Brandt et al., 2002] and IMAGE/LENA [Moore et al., 2000] were both launched within about the past decade, so the field, along with an understanding of the experimental techniques, is still maturing and, as discussed in the paper, the data presented here do not appear to suffer instrumental effects such as UV sources or direct solar wind
charged particle leakage.

\ \ \ Luckily, there are exciting current and future missions which will have neutral atom imaging capability, potentially contributing to answering or even resolving some of the questions that the LENA and Mars Express observations have raised: The BepiColombo mission to the Planet Mercury will include low energy neutral atom instrumentation and Solar Orbiter may have a dedicated neutral solar wind 
instrument [J. Hsieh and S. Orsini, private communication]. 
The Mercury Planetary Orbiter is anticipated to have a neutral atom imager. The Mercury Magnetospheric Orbiter will carry a low energy neutral atom sensor called ENA.
The Chandraayan-1 mission to the Moon will carry a copy of the ENA sensor.
Finally, the Venus Express mission carries ASPERA-4 which is a copy of the Mars Express instrument with both the NPI and NPD sensors on it.
Thus, we will eventually have data sets from similar instruments outside of 1 AU, at 1 AU, and inside of 1 AU for comparison.


\end{document}